\documentclass[a4paper,fleqn]{cas-dc2}

\usepackage{cas-dfrws} 
\renewcommand{\dfrwsconference}{Digital Forensics Research Conference Europe (DFRWS EU),  March 29--April 1, 2022}

\usepackage{stix} 

\usepackage[authoryear,longnamesfirst]{natbib}
\usepackage{booktabs}
\usepackage{fancyvrb}
\usepackage{arydshln}

\newenvironment{command}
    {\small%
    \ttfamily%
    \par%
    \vspace{1mm}%
    \$}
    {
    \par%
    \vspace{1mm}%
    }

\usepackage{xcolor}
\definecolor{ao}{rgb}{0.0, 0.5, 0.0}

\usepackage[colorlinks]{hyperref}
\hypersetup{
    linkcolor=CornflowerBlue,
    citecolor=CornflowerBlue,
    urlcolor=cyan
}

\newcommand{\urlDate}{last accessed 2021-02-22}
\newcommand{\furl}[1]{\footnote{\url{#1} (\urlDate).}}

\usepackage{verbatim}  

\begin{document}




\title[mode=title]{Identifying document similarity using a fast estimation of the Levenshtein Distance based on compression and signatures}
\shorttitle{Identifying document similarity using fast Levenshtein Distance estimation}

\shortauthors{P.~Coates \& F.~Breitinger}


\author[adr1]{Peter Coates} 
\ead{coatespt@gmail.com}
\ead[url]{https://hadoopoopadoop.com}

\author[adr2]{Frank Breitinger}[orcid=0000-0001-5261-4600]
\cormark[1]
\ead{frank.breitinger@unil.ch}
\ead[url]{https://FBreitinger.de}
\credit{Conceptualization, Writing - Review and Editing, Supervision}

\address[adr1]{Kingswood Road, Weehawken, NJ, 07086, United States}
\address[adr2]{School of Criminal Justice, University of Lausanne, 1015 Lausanne, Switzerland}

\cortext[1]{Corresponding author.}
\nonumnote{Copyright remains with the authors.}

\begin{abstract}
Identifying document similarity has many applications, e.g., source code analysis or plagiarism detection. However, identifying similarities is not trivial and can be time complex. For instance, the Levenshtein Distance is a common metric to define the similarity between two documents but has quadratic runtime which makes it impractical for large documents where large starts with a few hundred kilobytes. 
In this paper, we present a novel concept that allows estimating the Levenshtein Distance: the algorithm first compresses documents to signatures (similar to hash values) using a user-defined compression ratio. Signatures can then be compared against each other (some constrains apply) where the outcome is the estimated Levenshtein Distance. Our evaluation shows promising results in terms of runtime efficiency and accuracy. In addition, we introduce a significance score allowing examiners to set a threshold and identify related documents. 
\end{abstract}

\begin{keywords}
Levenshtein Distance \sep Edit Distance \sep Estimation \sep Document Similarity \sep Approximate String Matching \sep Fingerprint \sep Digest
\end{keywords}

\maketitle



\section{Introduction}
A crucial task for digital forensic investigations is the identification of relevant artifacts (evidence) where examiners often rely on automation to cope with the sheer amounts of data. One example is file classification, which is often done using hash functions: compute the hash value of a file and compare the hash against a whitelist (in case of a match the file can be ignored) or blacklist (in case of a match the file is regarded as potential evidence that requires further investigation).
As hash functions can only identify exact duplicates, they have been complemented by approximate matching (AM) a.k.a.~similarity hashing or fuzzy hashing \citep{breitinger2014approximate}. Instead of providing a binary decision, most AM algorithms provide a certainty score indicating their confidence that two artifacts are similar or related. 
For instance, on a scale of 0 to 100, where 0 indicates no similarity, a score of 80 indicates that the tool is confident that the two artifacts have similarity where the definition of similarity is defined by the tool and can be on various levels, e.g., byte-level similarity vs.~visual similarities in case of images.
While these algorithms provide great value to practitioners, there are two problems: 
First is the interpretation of the output as a score of 80 does not mean there is a similarity of 80 \emph{percent}. In fact, two artifacts can have a score of 100 but their cryptographic hashes are different. 
Second, AM algorithms work best for larger artifacts but are less reliable for medium sized texts, e.g., comparing spam Emails with each other to identify spam campaigns. Which, depending on the length of the text, is best solved using approximate string matching algorithms such as Levenshtein distance.

\paragraph{Levenshtein distance (LD)} LD is precise and has an intuitive meaning regarding the dissimilarity of two strings. Given a pair of strings, the LD is the number of single-character edits (i.e., insertions, deletions, or substitutions) that are required to turn one string into the other. For example, the LD of \texttt{pat} and \texttt{mat} is one (replace \texttt{m} with \texttt{p}); the LD of \texttt{pats} and \texttt{mat} is two (replace m with p and delete the s).  
The LD of pairs of randomly chosen longer texts of equal length tends to be narrowly distributed. Therefore, an observed divergence of the LD of a pair of texts from the expected mean can be interpreted mathematically and also has an intuitive meaning to a human.
Unfortunately, in practice, LD is primarily relevant for short strings (e.g., comparing two strings with 50'000 characters each requires already 3.5sec on a modern Laptop) because the time complexity of the algorithm is quadratic\footnote{It has been shown that LD cannot be calculated in better than quadratic time for the worst case, but it is possible to do better in special cases, for instance, for strings that are known to be similar. For simplicity, we ignore the special cases in this article.} 
being proportional to the product of the lengths of the two strings, i.e., $\mathcal({O}(|a|\times |b|))$. Quadratic performance does not present a hard size limit, but comparing larger strings, such as Web pages, long articles, or books, becomes impractically slow, and faster computers are of little help when confronted with a quadratic performance curve.

\paragraph{Contribution} The heuristic described in this article offers a partial solution to the problem by offering a way to \emph{estimate} the LD of texts pairs over a size range many hundreds of times larger than is practical for calculating an exact value. The algorithm works by first compressing the input (the result is called signature) and then by applying the Levenshtein Distance to the signatures rather than to the original texts. 
Note, as the heuristic described herein itself uses LD, it suffers from the same quadratic limitation. However, signatures can be typically many hundreds of times shorter than the original texts and therefore the algorithm produces estimates faster. 
For example: A size reduction factor of $C$ results in a speedup of approximately $C^{2}$. Viewed the other way around, if a given computation time is deemed acceptably fast for files of up to size $X$, the heuristic extends the tolerable size to $CX$ (where $C$ is a parameter and can be adjusted by a user).

\paragraph{Abbreviations} The following abbreviations are used

\begin{tabular}{ll}
    LD & Levenshtein Distance \\
    eLD & estimated Levenshtein Distance \\
    LCA & Lossy Compression Algorithm \\
\end{tabular}

For simplicity, we will use $LD$ and $eLD$ in the text as well as during the algorithmic description. Thus, $LD(A,B)$ means generating the Levenshtein distance between two documents $A$ and $B$.
Note that we differentiate between `signature' and `digest': a digest is the return value of our compression algorithm (similar to a hash value) where the signature includes the `digest' as well as other information (e.g., filename and file size).

\paragraph{Structure} The remainder of the paper is organized as follows: 
first, we describe our algorithm in Sec.~\ref{sec:the_algo}, which is the core contribution of this article, followed by a brief summary of the reference implementations in Sec.~\ref{sec:implementations}. Sec.~\ref{sec:eval_discussion} provides an assessment of our algorithm as well as a discussion. In Sec.~\ref{sec:df_application} we discuss some forensic applications from a high-level perspective and introduce the significance score allowing to filter for related documents. The last two sections are the \nameref{sec:rel_work} and the \nameref{sec:conclusion}.

\section{The Algorithm}\label{sec:the_algo}
We estimate (approximate) the LD of two documents by 
(1) compressing each document into a signature using a Lossy Compression Algorithm (LCA), 
(2) applying the LD algorithm to the compressed signatures, and 
(3) scaling the result back by the compression rate. 
As LD is well-defined, this chapter discusses compression algorithm in  Sec.~\ref{sec:compression_function}, the layout of the $eLD$ signature in Sec.~\ref{sec:eLD_signature} and the scaling / estimation of the LD in Sec.~\ref{sec:LD_estimation}.

\subsection{Compression Algorithms}\label{sec:compression_function}
As outlined in the subsequent paragraph, conventional compression algorithms such as LZ77 or BZIP2 are not suited to this purpose (Sec.~\ref{sec:conv_comp}). Given these limitations, we present properties necessary for our lossy compression algorithm in Sec.~\ref{sec:lossy_comp_prop} followed by the algorithmic description in Sec.~\ref{sec:comp_algo_des}.

\subsubsection{Conventional Compression}\label{sec:conv_comp}
One reason conventional compression algorithms are not suitable is that they perform `lossless' compression, which greatly limits their compression rate. Lossless compression is inherently limited by the amount of information present in the data to be compressed. In information theory terms, English text contains about 1.3 bits of information per 8-bit character, which makes typical text about 0.1625 information and 0.8375 `air' that can in principle be squeezed out without rendering the original irretrievable. This limits text compression to a factor of approximately 6:1.
A second reason is that we want the compressed version to be something like the thumbnail version of a photograph. This means that despite being many times smaller (in size) than the original, a signature must preserve some kind of recognizable similarity to the original if direct comparison of signatures is to provide a meaningful result. Ordinary compression algorithms systematically eliminate this kind of local similarity because maximum compression means that the result looks uniformly random.

\subsubsection{Properties for a lossy compression algorithm}\label{sec:lossy_comp_prop}
When estimating the LD, we rarely if ever are concerned with rates as low as the highest rates conventional compression can achieve. A compression rate of 25:1 would be quite small for an estimation application; compression factors from 100 to the low thousands are more usual for text and even larger compression factors may be used for binary data such as videos.

For our approach, the compression function requires the following properties (note: the term (lossy compression) \emph{digest} is used to describe the output of our compression function):
\begin{itemize}
    \item Compression: The digest must be much smaller than the original input where `much' is a factor of 100 to the low thousands.
    \item Determinism: The digest must be identical for identical inputs.
    \item Runtime efficiency: Ideally, the compression algorithm runs in linear time with respect to the length of the input. At least, it has to be fast.
    \item Concatenation: The result of concatenating the digests of two strings shall be identical to concatenating the two strings first and then calculating the digest, i.e., $dig(A)+dig(B) = dig(A+B)$ .
\end{itemize}

Our proposed algorithm does not fully satisfy the fourth property and therefore we settle for ``good enough.'' That is, if we concatenate strings first and then generate a digests ($dig(A+B)$), the difference from the concatenated digests ($dig(A)+dig(B)$) must be limited to differences arising from a bounded number of characters on either side of the point of concatenation of the originals. 
As we will elaborate on below, the number of characters around the point where strings abut that can affect the output is determined by one of the two key compression parameters.

\subsubsection{Lossy Compression Algorithm (LCA)}\label{sec:comp_algo_des}
\begin{figure}[th]
    \centering
    \caption{A simple visualization of our compression algorithm; at $p=3$ the sequence \texttt{LLO\_WOR} is compressed to the letter `a' (hash values are randomly chosen and do not match the actual implementation).}
    \label{fig:compression_algo_overivew}
    \includegraphics[width=0.48\textwidth, trim={4cm 3.5cm 5cm 0.5cm}, clip]{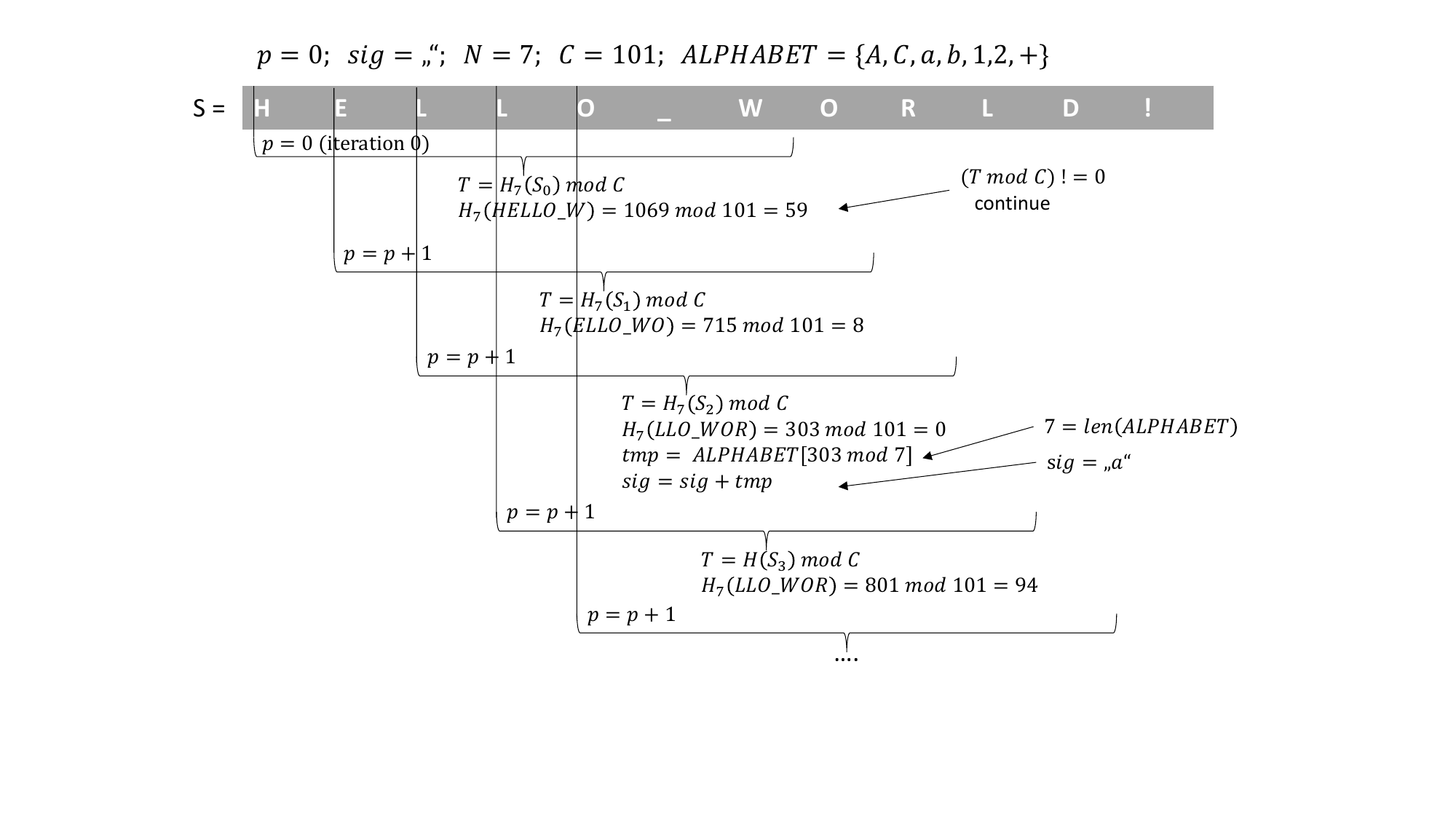}
\end{figure}

Our LCA uses a rolling window of size-$N$ (neighborhood) that slides through the text (character by character) and generates a pseudo-random value (hash) at each position (if the document has $L$ characters, $L-N+1$ neighborhoods are processed). 
Depending on a neighborhood's hash, our LCA either 
(a) does nothing, which is by far the most common decision or 
(b) uses the hashed value to select a single character from a fixed set `replacing' the neighborhood in the digest.
This keyhole view of the data ensures that any substring of the input can have only a localized effect on the output. 
The exact procedure is as follows:

\paragraph{LCA description} Let $S$ be the input of length $L$, and let $S_p$ denote the current position $p$ in $S$ with $0 \le p \le L-N+1$. 
$H_N(S_p)$ denotes the hash function on a substring of length $N$ starting at $p$ in $S$ (details about $H$ are discussed in Sec.~\ref{sec:LCA_hash}).
Let $ALPHABET$ be an array of unique characters (e.g., [a...z,A...Z,0...9]). Then our LCA works as follows (a visual description of the implementation is provided in Fig.~\ref{fig:compression_algo_overivew}):
\begin{enumerate}
    \item $dig$ (lossy compression digest) denotes a string\footnote{String was chosen for simplicity and readability; any other format such as binary, hex or base64 would be possible and only requires minor changes.} containing the digest, which is initially empty. The compression rate $C$ is an integer and discussed later.
    \item If $p+N \ge L$, print $dig$ and quit ($p$ is initially zero).
    \item Compute $H_N(S_p)$ and store the result in $T$.
    \item If $(T ~mod~C) ==0$\footnote{While we chose 0, any other arbitrarily chosen value between $0$ and $C-1$ will work.},  
        \begin{enumerate}
            \item generate $tmp = T~mod~len(ALPHABET)$, and
            \item append $ALPHABET[tmp]$ to $dig$.
        \end{enumerate}
    \item Set $p=p+1$ and return to step 2.
\end{enumerate}

\noindent The three required parameters ($C, N, ALPHABET$) impact the algorithm as follows:
\begin{itemize}\setlength{\itemsep}{0pt}
    \item The nominal compression rate $C$: 
    The larger $C$, the less likely the if-statement (step 4) is triggered and thus the smaller the digest will be. The choice of $C$ is subject to the following limitations: 
    if $C$ is too large, too much information is lost and the digest provides little value (or may even be an empty string); 
    if $C$ is too small, digests will be bulky requiring more storage and comparing digests using LD will be slow.
    The optimal choice depends on the use case. 
    \item The neighborhood size $N$: Compression operates on size-$N$ substrings of the input, one neighborhood starting with each successive character. The larger $N$, the more sensitive the heuristic will be to small differences because each character is part of the $N$ distinct neighborhoods (except within $N$ characters of the beginning or end of the input).  
    For instance, let us assume an input with $L=1000$ and $N=500$. Performing two changes, e.g., at positions $p=\{300,700\}$, will likely change the resulting digest completely.
    On the other hand, if $N$ is too small, the lack of diversity in the neighborhood substrings results in an uneven distribution of output characters. Common values for $N$ range from 11 to 21.
    
    \item The $ALPHABET$ describes the characters used to build the digests, e.g., printable ASCII characters except newline, tab or space, and are chosen to be conveniently readable for the user. The larger the $ALPHABET$, the less likely two different $T$'s are mapped to the same character. Note, the length of $ALPHABET$ has to be mutually prime to $C$, (for simplicity one may use an $ALPHABET$ of length prime). If they are not mutually prime, different strings will be mapped to the same output character more likely.
    $ALPHABET$ cannot include `,' (comma) as it is used as a separator in the final signature.
\end{itemize}

\noindent Given the $ALPHABET = [a...z,A...Z,0...9]$, a digest generated by the LCA may look like: \texttt{AeVVgCAe2aZUpa6dnEkK...}
    
\subsubsection{Selection of Neighborhood Hash Function}\label{sec:LCA_hash}
As every $N$-size neighborhood is hashed resulting in $L-N+1$ calls of the hash function $H$, its runtime efficiency is crucial. Naturally, a rolling hash function (e.g., Rabin Karb) or a non-cryptographic hash function (e.g., djb2) seem to be a good choice as they are performant. 
To analyze the impact of $H$, several different hash functions were tested and the results are summarized in Sec.~\ref{sec:gen_runtime_eff}. Our final implementation uses Rabin Karb fingerprint but the source code allows to easily replace this function.
Note, Rabin Karb is a non-cryptographic hash function and does not have crypto properties such as preimage resistance or collision resistance. However, the lossy compression ensures that the preimage cannot be recovered but it is possible to generate an input that produces a given digest.

\subsection{eLD signature}\label{sec:eLD_signature}
The final $eLD$ signature comprises of a \emph{header} as well as the LCA \emph{digest} (in the following we will use the term digest instead of LCA digest). The header is necessary, as only a comparison of signatures with identical parameters is possible. Furthermore, additional information is required to do the scaling as well as assessing the quality of the potential match. The final $eLD$ signature includes the following values separated with a comma (* indicates mandatory values): 

\begin{itemize}
    \item The path/filename of the original file.
    \item The length of the original file*.
    \item The $C$ value used to generate the signature.
    \item The $N$ value used to generate the signature.
    \item The length of the LCA digest 
    \item The LCA digest*.
\end{itemize}

Values that are not essential can be suppressed in the output, e.g., if $C$ and $N$ are immutable they are not required in the signature. A sample output is given in Fig.~\ref{fig:sample_sig}

\begin{figure}[th]
\caption{Sample $eLD$ signature output including all six values; the signature has been shortened for better readability; comment line (\#) was added manually for readability and is usually not printed.}\label{fig:sample_sig}
\scriptsize
\begin{Verbatim}[frame=single]
$ go run *.go -in test-data/doc.txt 
#filename, fileLength, C, N, digestLength, digest

test-data/doc.txt,13680,101,20,152,7n(n[(9&dU7;aRZaPgSGWzoFCC_r
{FUB;A]v?dGrL.bQZ!3GCT)V0r>XWpPNmQ6>#8YhTN]h-X598u+qw1Q1K:+&CVI
gIhCK//j2 ...
\end{Verbatim}
\end{figure}


\subsection{Estimating the LD}\label{sec:LD_estimation}
To estimate the LD of two documents given their $eLD$ signatures, we first compute the LD between the digests and then perform a scaling where ideally $LD(A,B) \approx eLD(A,B)$. 
Before explaining the estimation, it is essential to realize that there are two scenarios:

\begin{enumerate}
    \item The documents have roughly the same size and therefore the generated digests shall also have approximately the same size. 
    \item The documents have a large difference in size (e.g., ratio 1:5) which also results in a large difference of the digests length. 
\end{enumerate}

In addition, we have to consider that two unrelated documents still have a LD shorter than their length. Meaning that, if $A$ and $B$ are two texts with 1000 characters each, then the maximum $LD(A,B)$ would be 1000 but usually it is less as there is an overlap by chance. This phenomenon will be called \emph{expected overlap} $R$.
As a consequence, it requires more than only multiplying the result by the compression ratio. 

\paragraph{Expected Overlap Ratio $R$} The expected overlap ratio $R$ is mostly influenced by the underlying alphabet and the letter frequency. For instance, the alphabet in an English book is almost all 100 printable ASCII characters\footnote{Python \texttt{\$ len(string.printable)}} but the distribution and order is not random, e.g., there will be spaces between words, letter `e' is most frequently used, `the' is a frequent word, etc.
To obtain $R$, we programmed a python script\footnote{\texttt{getR.py} in the helper-directory.} that, given an alphabet, creates two test-strings ($TS$) of length 30'000 using randomly chosen alphabet elements and calculates the LD. The three alphabets are:
\begin{description}
    \item[random\_chars:] This alphabet consists of 83 elements \\ \texttt{string.ascii\_letters}, \texttt{string.digits} and \\ \texttt{()[]+\#\_-!?\%<>@.:;\&/\{\}}.
    \item[eng\_chars:] For this alphabet we parsed a book, removed newlines and generated one large string with 686'250 elements. As we randomly select one element at a time, this reflects the frequency of letters in English texts, i.e., the $TS$ has many spaces and `e' but only a few `!' or `X'.
    \item[eng\_words:] Similar to before, we parsed a book but converted it into a list containing words instead of one large string. Thus, the alphabet is a list of 97'591 words. When creating $TS$, we added a space after each word. 
\end{description}
Example $TS$ for each alphabet are provided in Fig.~\ref{fig:test-strings}. The estimated overlap ratio $R$ is then calculated as follows:
\begin{equation}
    R_{alphabet} = 1 - \frac{LD(TS_1, TS_2)}{len(TS_1)}
\end{equation}
where $len(TS_1) = len(TS_2) = 30'000$. For each $R_{alphabet}$, we performed 10 runs and averaged the ratio resulting in: $R_{random} = 0.0417$, $R_{chars} = 0.1593$ and $R_{words} = 0.1902$. 

\begin{figure*}[th]
\caption{Shortened examples of the test-strings used to calculate $R$ based on the different alphabets.}\label{fig:test-strings}
\footnotesize
\begin{Verbatim}[frame=single]
random_chars: 9SRCv8A{CKO&h{Ly0u7Nu)gw0(OE(0e/!AMm#ss-IGu&CIAoB@lCbwF#Rn:/(Br5GC.6U??iMe.]]t{/VQcd#jW0sTpS?+Xi8GKO0uNVf(6qX50f[2TA&z##X1.

eng_chars: stwbeulahf yafyes tagokku  seuca  niuaasdnnicollneptdCannags t gahoni hiyneeet ntdtngn eio;awgptmpao RrD eu1lc;ccdsLnPtrs7r il

eng_words: are Penn still Logan? the John of compensation Nicola sins Wilcox,along booklet on 9, Louis for spur trees, Caleb of in volume
\end{Verbatim}
\end{figure*}

\paragraph{eLD scaling} Let $A$ and $B$ be two documents with $len(A) \ge len(B)$ (we expect the same for the digests, i.e., $len(dig_A) \ge len(dig_B)$).
Then the estimated LD is calculated as follows (for readability $|~|$ replaces the $len$ function):

\begin{equation}\label{eq:eld}
\footnotesize
\begin{split}
    \texttt{digDiff}        & =   |dig_A| - |dig_B| \\
    \texttt{effectiveC}     & =   ( |A| + |B| ) / ( |dig_A| + |dig_B| ) \\
    \texttt{digLD}          & =   LD( dig_A, dig_B ) \\
    \texttt{scaledDigLD}    & =   \texttt{( digLD - digDiff ) * effectiveC} / ( 1 + R_{words} )  \\
    \texttt{fileLengthDiff} & =   |A| - |B| \\
    eLD                     & =   \texttt{round( scaledDigLD + fileLengthDiff )}
 \end{split}
\end{equation}

That means: $eLD$ consists of the difference in document length \texttt{fileLengthDiff} as for every missing character, one LD operation is neede (the document lengths ($|A|$, $|B|$) are stored as part of the signatures).
The second part of the estimation, \texttt{scaledDigLD}, which has to be added to \texttt{fileLengthDiff}, focuses on the portion of the digests that do \emph{not} match, i.e., $dig_A = 12ABCD$ and $dig_B=12FE$, this focuses on the $FE$ (\texttt{digLD - digDiff}); the $12$-part can be ignored as the LD=0. 
The effective compression rate \texttt{effectiveC} is used for scaling. Lastly, we know that there is an expected overlap which we factor in by multiplying ($1 + R_{words}$).

\paragraph{Sample calculation} Let us assume two documents (700 and 500 bytes) that have the first 260 bytes in common. They have the following signatures ($C$ and $N$ are identical thus signatures can be compared):\\
\verb+   #filename,fileLength,C,N,digLength,digest+\\
\verb+   docA,700,51,20,15,AABBCFF00192192+\\
\verb+   docB,500,51,20,10,AABBCCDDEE+

\noindent We will now follow Eq.~\ref{eq:eld} and calculate the eLD:
\begin{enumerate}
    \item digDiff = 15 - 10 = 5 (to get the length of the digests we can either count or use the digLength-column
    \item effectiveC = ( 700 + 500 ) / ( 15 + 10 ) = 48 (we had a compression rate of $C=51$ but the actual compression is likely different so we compute the actual compression rate).
    \item digLD = (AABBCFF00192192, AABBCCDDEE) = 10 (replace CDDEE with FF001 and add 92192).
    \item scaledDigLD = ( 10 - 5 ) * 48 = 240 (the five characters CDDEE corresponds to half the signature and should roughly cause an LD of 250); however docA and docB have likely some overlap in this region wherefore we reduce 250  / ( 1 + 0.19 ) = 201.68 
    \item fileLengthDiff = 700 - 500 = 200 
    \item eLD = round( 200 + 201.68 ) = 402
\end{enumerate}

If we compare $eLD$ to our starting situation (260 bytes in common), we know that the LD must be between $200 \le LD(docA, docB) \le 440$ where 200 is impossible as this would mean they have the first 500 bytes in common which is a contradiction.

\section{Reference Implementation(s)}\label{sec:implementations}
There are three reference implementations of the algorithm available - Java 11, Python 3.9 and GoLang 1.17 (thus all are platform independent). As they were developed and modified in parallel by different developers, they all work slightly different but were used to validate results. In the following we will focus on the GoLang implementation; details about the Java implementation can  be found in the \ref{app:eLD_java}; the Python implementation turned out to be significantly slower and will not be discussed further. All implementations are open-source and available online\footnote{The GoLang version can be downloaded here: \url{https://www.fbreitinger.de/?page_id=218}; the Java implementation is available here: \url{https://coatespt.github.io/Fast-Levenshtein-Distance/}; for Python please contact the authors as it is not maintained.}.

The zip file contains 
our go implementation in the \texttt{eLD} folder, some test data, as well as an overview of the test results. Options can be found in \texttt{eLD/config/eLDconfig.go} To generate signatures for the files-folder run
\begin{command}
go run *.go -in /path/to/dir > signatures.csv
\end{command}
\noindent and to compare all signatures within a signature file run
\begin{command}
go run *.go -source signatures.csv
\end{command}
The help menu laying out the other options can be access using the \texttt{-h} option.

Remark: As it is an early-stage prototype, error handling and logging is currently limited. It may also be possible to optimize code and make the implementations more efficient.

\section{Evaluation}
To assess our implementation, we ran experiments looking into 
\nameref{sec:gen_runtime_eff} (Sec.~\ref{sec:gen_runtime_eff}), 
\nameref{sec:cmp_runtime_eff} (Sec.~\ref{sec:cmp_runtime_eff}), 
\nameref{sec:eLDvsLD} (Sec.~\ref{sec:eLDvsLD}), and 
\nameref{sec:eLDvsLD_rel} (Sec.~\ref{sec:eLDvsLD_rel}). 
The last section is a \nameref{sec:eval_discussion} (Sec.~\ref{sec:eval_discussion}).
All experiments were conducted using the GoLang reference implementation summarized in the previous section.

\paragraph{Device} Test results were obtained on a MacBook Pro running BigSur version 11.06. The system used for testing has an 8-Core Intel Core i9 @ 2.3\,GHz with 64\,GB Memory and an Apple SSD AP1024N 1TB drive. During all tests, the system was in use meaning that other processes like Browser, Email client, etc. were running. 

\paragraph{Dataset} As our implementation is geared towards text files, we utilize the Gutenberg project and downloaded 18,216 files ranging from 10\,KB to 12\,MB totaling in approximately 7.2\,GB. Subsets were created as needed and are explained in the corresponding section.

\subsection{Signature generation runtime efficiency}\label{sec:gen_runtime_eff}
The efficiency of generating signatures is dominated by the speed of the hashing algorithm as it has to process all neighborhoods. 
To test the performance of different hashing algorithms, we created four versions of our tool utilizing Rabin Karb, djb2, FNV and MD5. 
For each setting ($N=\{7,14,21\}$), we did three runs and averaged times. The \texttt{time} command was used for measurement and we provide the \texttt{user+sys} time which denotes the actual CPU time of the process\footnote{\texttt{\$ go run} will compile and run the program; compiling first using \texttt{\$ go build} and then running the application would be slightly faster.}:
\begin{command}
time go run *.go -in ../allfiles/ > res.db
\end{command}
The results are shown in Table~\ref{tab:gen_speed}; as a Benchmark we also processed the dataset with \texttt{shasum} which required 14s.

\begin{table}[th]
\caption{Average runtime for different hashing algorithms using three different $N$ and a constant $C=301$.\vspace{2mm}}\label{tab:gen_speed}
\centering
\begin{tabular}{ l r r r }
\toprule
    N                   & 7		    & 14	        & 21	    \\ 
\midrule
    Rabin Karb	        & 42s       & 43s 	        & 42s	    \\
    djb2            	& 1m\,55s 	& 2m\,35s       & 3m\,18s   \\
    FNV*     	        & 2m\,13s 	& 2m\,39s 	    & 3m\,21s 	\\
    MD5*     	        & 19m\,18s  & 19m\,41s 	    & 19m\,39s 	\\
\midrule
\multicolumn{4}{l}{\parbox{6.5cm}{\footnotesize *Tests for FNV and MD5 utilized GoLang libraries and required type-casts which may have impacted runtime.}} \\ 
\end{tabular}
\end{table}

As expected, $N$ does have no impact on Rabin Karb fingerprint but on other algorithms that process sequences and not char-by-char. It also has no impact on MD5 as long as $N$ is smaller than the MD5 block size of 512\,bit. A constant $C$ was utilized since it has little/no impact on the runtime: $C$ influences the digest length (we confirmed our assumption by running some tests using different $C$).
Lastly, we also ran some tests with the Java as well as Python implementation but both were significantly slower and will not be discussed here, e.g., Java using the \texttt{string.hashcode()} required usually around 5 minutes.

\subsection{Signature comparison runtime efficiency}\label{sec:cmp_runtime_eff}
While $C$ has no impact on the signature generation, it influences the signature comparisons. Remember, the larger $C$, the smaller the final digest and the digest comparison uses LD with $O(L^2)$. 
For this test, we randomly selected 40 files (between 1.2\,MB and 22\,KB) from our dataset with a total of 17.4\,MB and created a signature file. We then measured the time for an all-against-all comparison (i.e., 780 comparisons total) with various $C$'s. Times and signature file sizes are summarized in Table~\ref{tab:spd_by_C}. 

\begin{table}[th]
\caption{Efficiency for comparing the sample signatures for various $C$ $(N=11)$; csv size is the file size of the signature file used for comparison.}\label{tab:spd_by_C}
\centering
\begin{tabular}{ l r r }
\toprule
\bf C       & \bf duration      & \bf csv size  \\
\midrule
51      & 2m\,12s       & 342\,KB   \\
101     & 34s           & 175\,KB   \\
301     & 4.2s          & 59\,KB    \\
501     & 2.0s          & 37\,KB    \\
1001    & 0.9s          & 19\,KB    \\
\bottomrule
\end{tabular}
\end{table}

\subsection{Estimated LD vs.~LD (unrelated/loosely related files)}\label{sec:eLDvsLD}
This section compares our estimated LD vs.~the original LD with respect to efficiency and accuracy for unrelated/loosely related files (as all documents originate from the Gutenberg project, they share common header and footer information which we did not remove for this test). 
We created another subset using 20 random files from the original dataset where each document was required to be between 20\,KB and 40\,KB ensuring that LD completes in a reasonable time. In total, the 20 files had 627\,KB. An overview of the experiment results is given Table~\ref{tab:eLD_vs_LD} where the LD-column shows the original LD followed by various $C$'s. Our test ignored self-comparisons, i.e., docA is only compared against other documents and not itself, resulting in a total of 190 comparisons. 

\begin{table*}[th]
\caption[na]{$LD$-column shows the time for a traditional LD implementation from \cite{GolangLD}; the remaining columns outline $eLD$ results for various compression rates $C$ where $N=11$ and \texttt{EXPECTED\_OVERLAP\_TEXT=0.19} remained unchanged. Columns provide the absolute value and the percentage value (rounded to one digit).}\label{tab:eLD_vs_LD}
\centering\small
\begin{tabular}{ l c r r r r r }
\toprule
                                        & $LD$	        & $C=11$	        & $C=21$	        & $C=51$	        & $C=101$	        & $C=201$               \\ 
\midrule
    Duration	                        & 9min\,09s     & 3.5s              & 1.2s              & 0.6s              & 0.4s              & 0.3s              \\
    Avg Error (abs/\%)	                & -- 	        & 1223 / 6.5\%      & 1166 / 6.4\%      & 1519 / 9.0\%      & 1578 / 9.0\%      & 1650 / 9.4\%      \\
    Min Error (abs/\%)	                & -- 	        & 16 / 0.1\%        & 10 / 0.1\%        & 10 / 0.1\%        & 3 / 0.0\%         & 16 / 0.1\%        \\
    Max Error (abs/\%)	                & -- 	        & 3498 / 23.2\%     & 3484 / 23.1\%     & 5177 / 34.3\%     & 4187 / 40.7\%     & 6942 / 35.6\%       \\
    Std Dev.  (abs/\%)                  & --            & 822 / 3.6\%       & 767 / 3.8\%       & 1114 / 6.4\%      & 1075 / 7.0\%      & 1424 / 7.3\%       \\
    Error rate (avg/std dev)            &               & 0.03 / 0.02       & 0.03 / 0.02       & 0.04 / 0.03       & 0.04 / 0.02       & 0.05 / 0.04       \\
\bottomrule
\end{tabular}
\end{table*}

To determine the avg error (row 2), we summed all absolute values of our estimate LD minus the LD and divided it by 190. In addition, we provide the min error, max error as well as the standard deviation. 
This raw error does not consider the ratio of the LD to the file length which is necessary to interpret the meaning of the results. 
For instance, let us assume pairs documents (size 32'000) having LD 1, 1000 and 15'000 and their eLD are 2, 2000 and 30'000. In all cases the relative error ($\frac{eLD-LD}{eLD}$) is 50\% but the impact is different: 
LD 1 / eLD 2 both indicate a strong correlation between the files while LD 15'000 indicates correlation but eLD 30'000 indicates unrelated documents. Therefore, we also put the error rates in perspective (we use the error rate as defined by \citet[Eq.~1]{mackenzie2002character} but we then deduct the rates to obtain a  single value):
\begin{equation}
    ER = abs(\frac{LD(A,B)}{max(|A|, |B|)} - \frac{eLD(A,B)}{max(|A|, |B|)}) .
\end{equation}
Consequently, if LD and eLD have similar error rates (which is desired), $ER$ is low. For our example, this would result in the following error rates: ($abs(\frac{1}{32'000}-\frac{2}{32'000})=$) 0.00, 0.03 and 0.47 , respectively. 

As shown by our results, our algorithm is significantly faster and results have a low `ER'. The larger $C$ the less accurate is the algorithm which may also be due to the rather small sample files (e.g., the average digest length is 158). Detailed results are also included in the downloadable zip file.

\subsection{Estimated LD vs.~LD (related files)}\label{sec:eLDvsLD_rel}
For this second test we reused the 20-random-file dataset from the previous section, created a modified copy of the file and then compared the original version against the modified version. A summary of the outcome is provided in Table~\ref{tab:eLD_vs_LD_rel}.

\begin{table*}[th]
\caption{Results of comparing one document against a modified version of itself manipulated according to the operation-column(modifications were done manually and randomly). FS and LD represent the original file size and the Levenshtein distance. The remaining columns show the $eLD$ for various $C$ (again $N=11$ and \texttt{EXPECTED\_OVERLAP\_TEXT=0.19} remained unchanged).}\label{tab:eLD_vs_LD_rel}
\centering
\resizebox{2\columnwidth}{!}{%
\begin{tabular}{rlllrrrrrr}
\toprule
 \#  &             &                & \textbf{Operation}                    & \textbf{org. FS}     & \textbf{LD} & \textbf{C=11} & \textbf{C=21} & \textbf{C=51} & \textbf{C=101} \\
\midrule
 1 &   10348.txt   & 10348mod.txt   & Deleted 10 lines                             & 24226         & 663         & 718           & 798           & 709           & 663            \\
 2 &   ltplt10.txt & ltplt10mod.txt & Deleted 10 lines                             & 22869         & 634         & 672           & 652           & 717           & 719            \\
 3 &   7563.txt    & 7563mod.txt    & Deleted 51 lines in the beginning            & 23414         & 1024        & 1024          & 1024          & 1024          & 1024           \\
 4 &   7565.txt    & 7565mod.txt    & Deleted 101 lines in the middle              & 25983         & 2234        & 2234          & 2234          & 2234          & 2330           \\
 5 &   7906.txt    & 7906mod.txt    & Deleted 90 lines (some beginning; some end)  & 21586         & 3329        & 3329          & 3329          & 3329          & 3329           \\
 6 &   10630.txt   & 10630mod.txt   & Deleted 7 chunks                             & 32185         & 7085        & 7095          & 7103          & 7086          & 7086           \\
 7 &   17282.txt   & 17282mod.txt   & Deleted 3 large chunks                       & 25534         & 7436        & 7436          & 7436          & 7436          & 7436           \\
 8 &   18232.txt   & 18232mod.txt   & Deleted 15 times 3 lines                     & 31635         & 2964        & 3117          & 3022          & 3034          & 3165           \\
 9 &   8528.txt    & 8528mod.txt    & Deleted approx first half                    & 33625         & 15811       & 15811         & 15811         & 15811         & 15811          \\
10 &   11592.txt   & 11592mod.txt   & Deleted 10 paragraphs                        & 36253         & 4224        & 4262          & 4259          & 4224          & 4224           \\
11 &   16637.txt   & 16637mod.txt   & Deleted 5 paragraphs at the beginning        & 39864         & 1731        & 1740          & 1731          & 1731          & 1731           \\
12 &   wlett10.txt & wlett10mod.txt & Deleted big middle chunk                     & 32346         & 8968        & 8968          & 8968          & 8968          & 8968           \\
13 &   lf17w10.txt & lf17w10mod.txt & Deleted all (450) `the'                      & 38902         & 1215        & 5409          & 5749          & 7128          & 7156           \\
14 &   haw4610.txt & haw4610mod.txt & Inserted 10 random `A'                       & 31340         & 10          & 93            & 113           & 96            & 10             \\
15 &   haw7810.txt & haw7810mod.txt & Inserted 40 random `A'                       & 36281         & 40          & 531           & 562           & 819           & 390            \\
16 &   lf20w10.txt & lf20w10mod.txt & Swapped 6 paragraphs around                  & 37765         & 3218        & 2831          & 2969          & 2897          & 1559           \\
17 &   14814.txt   & 14814mod.txt   & Swapped first and second half (approx)       & 28050         & 17517       & 14128         & 14454         & 15134         & 17721          \\
18 &   12337.txt   & 12337mod.txt   & All `b' (338) replaced with `B'              & 31241         & 338         & 4204          & 3172          & 3004          & 3248           \\
19 &   17195.txt   & 17195mod.txt   & All `e' (3080) replaced with `E'             & 35158         & 3080        & 21629         & 21605         & 22510         & 20106          \\
20 &   13322.txt   & 13322mod.txt   & All spaces (6572) replaced with double-space & 38838         & 6572        & 34083         & 35883         & 31128         & 32218          \\

\bottomrule
\end{tabular}
}
\end{table*}

The proposed estimation works well for deletions as long as they are not spread out across the complete document (compare L1-12 vs.~L13). Similarity, random insertions (L14+15) have a similar impact on the performance although if there are only a few insertions and a higher compression rate, this may be equalized. Swapping (L16+17) is also handled well where it is eye-catching that the estimated distances are lower than the original distance. Clearly, our algorithm has difficulties if there are many minor changes throughout the document (L18-20). However, these may be reduced by some additional pre-processing such as trimming / stripping or making it case insensitive.

\subsection{Discussion of the evaluation}\label{sec:eval_discussion}
The initial assessment  primarily focused on the $C$-parameter and runtime; more testing is required to assess the influence of other parameters such as $N$ or $R$ (\texttt{EXPECTED\_OVERLAP\_TEXT}). However, this preliminary evaluation shows promising results and probably can be optimized by adjusting other parameters.
Currently, our approach performs well as long as the files are of sufficient size to generate adequate signatures -- for small files one can fall back to the original LD.

Estimating signatures has both advantages and limitations: Most obviously, if the application at hand requires a precise LD value, the proposed heuristic is not useful. For instance, the heuristic will often estimate an LD of zero for texts that in fact vary slightly. It will never, however, do the opposite, i.e., estimate that two documents have an LD greater than zero when in fact they are identical.

Algorithmic performance depends upon specific signature-generation parameters as well as the text itself but estimates are typically within a few percent of the LD and tend to be more accurate with relatively similar documents, which is often the most useful category. Meaning that there are more efficient ways to identify very/completely different documents such as approximate matching (see Sec.~\ref{sec:rel_work}).
The level of accuracy is sufficient for tasks like detecting near-duplication and partial duplication, estimating how much two documents vary, filtering through large numbers of documents to look for a near-match to some substantial block of text, and detecting documents that are related in more complex ways. 

The effectiveness also depends upon variability in the input; there can be inputs that produce pathological results. For instance, the same short (relative to $N$) character sequence endlessly repeated may result in a `zero' length output or in an excessive output because one or more of the $N$-length substrings that are repetitively encountered causes an output character to be generated. 
While repetitive input of any great length would be unusual in natural language text, it can show up in various kinds of machine-generated text such as CSV files, log files, experimental data, etc. 
From a forensic perspective,  signatures for such data are unlikely to be useful (i.e., one would probably not compare them against other documents).
If excessively short or empty, they would fail to match even nearly identical files, and if they were excessively long, they would result in excessive processing time. For realistic $N$ and $C$, neither of these results will be close to the expected output size of $\frac{1}{C}$ of the input size and thus can be easily identified.

The signature of a fragment of a document is identical with the corresponding portion of the signature of a larger document in which it is embedded (except possibly near the points where they abut.) Therefore, signatures of related documents differ in locations that are in approximately the same relative positions as the corresponding differences in the originals. This means, not only can signatures be used to estimate how much two documents differ, they can also be used to infer useful information about where and how they differ. For example, they can indicate that two documents differ mostly at the beginning and end but differ little elsewhere, or that many differences are sprinkled throughout, or that one document appears to be appended to the other.

One of the most useful aspects of the signature-based approach is that neither the target documents nor the search documents need be in hand at the time an estimate is made. This has the obvious benefit of allowing the cost of signature generation to be amortized over many uses, but it also means that it is possible to search a collection of signatures for similar documents and assess their degree of textual similarity of candidates without having direct access to any of the documents involved.

\section{Digital forensic application}\label{sec:df_application}
While many files on computers are binary, there is still plenty of text based information that must be assessed during an investigation which is the application of our prototype implementation.
Having an efficient implementation of the Levenshtein distance allows the algorithm not only to be applied on short sequences but also larger files such as complete documents or source code files to cluster files, detect plagiarism or utilize it for black-/whitelisting. In the following we outline four scenarios where our algorithm can contribute:

\begin{description}
    \item[Source code analysis:] During an investigation it may be necessary to analyze and compare large software projects in order to identify if there are similarities in the source code. Identified similarities could mean that source code was copied (stolen) and reused. On the other hand, this also allows clustering similar files together. An example here would be the analysis of malicious code where an investigator received access to the source code of thousands of malware samples and wants to cluster similar samples together reducing the time of the manual analysis (e.g., we assume that similar malware applications have a similar behavior and only one sample per family, i.e., cluster, needs to be analyzed).
    
    \item[Fuzzy string matching:] Instead of looking for exact matches, it may be necessary to identify strings that roughly match a given string or are a substring of a given string. For instance, let us assume an examiner is given a Spam-Email and has to proof that the suspect is responsible for an Email spam campaign wherefore s/he decides to search in all documents for the given Email. It is likely that the Email has been modified throughout the campaign to avoid detection by spam filters.

    \item[Similarities between cybercrimes:] Besides processing evidence, our approach may also be useful to identify similarities between cybercrimes as discussed by \cite{bolle2018using}. In their work, the authors propose ``new approach to finding linkages and repetitions across cases in a cyber-investigation context using near similarity calculation of distinctive digital traces''. In their study, the authors use the Levenshtein distance on email addresses. Using a more performant algorithm allows to not only analyze email addresses but to cross compare additional (more comprehensive) traces.
    
    \item[Anonymous searches:] As little useful information about the contents of documents can be divined from the signatures, one can have a service that allows an outside party to check whether partial matches to a document or other files are present without the service having access to the library to be searched or to the text to be searched for. The client does not have to provide the query and the service only has access to the signatures, not the originals.
\end{description}

Note, while the current prototype has been developed for text files only, we plan  to expand it and see if it can be used for binary files as well.

\subsection{Significance score}
The estimated LD is computed for each pair of signatures, but in most cases, an estimate is not informative without context. 
To assess the importance of a comparison and filter out pairs that are not in a relationship (which is hard to do from the estimated LD score), we introduce the \emph{significance} $\delta$\footnote{This is work in progress and as will be seen the current score has some weaknesses requiring a workaround.}. 
This value puts the $eLD$ in relation to the digest lengths and allows an examiner to set a threshold $T$ ($0.0 \le T \le 1.0$) so that only matches with $\delta \ge T$ are shown. Note, $T=0.0$ will return all matches and $T=1.0$ will only return `exact' matches with $eLD=0$.

As discussed in Sec.~\ref{sec:LD_estimation}, there are two different scenarios where digests have a similar length or may vary widely. Ideally, $\delta$ considers both cases. That is, $delta$ shall be high if a smaller digest is (almost) included in a larger sequence as well as it shall be large if there is a low LD between two similar sized digests. Currently, the significance score $\delta$ for two digests is computed as follows (we assume $len(dig_A) \ge len(dig_B)$):

\begin{equation*}\label{eq:sig_score}
\begin{aligned}
        ld  & = LD(dig_A, dig_B)                    && \text{Levenshtein distance of digests} \\
    \delta   &= \frac{len(dig_A)-ld}{len(dig_B)}
\end{aligned}
\end{equation*}

However, this calculation has the weakness that it does not work for very large differences in digest size (e.g., ratios of 1 to 20). Table~\ref{tab:delta_values} provides some sample $delta$ values for given digests length and overlap where the top third contains `good' matches, the middle third contains `poor matches' and the last third demonstrates where $\delta$ fails.
The problem is, if the difference is too large, $dig_A$ can be converted to $dig_B$ just by deleting characters resulting in $LD(dig_A, dig_B)=len(dig_A)-len(dig_B)$.

\begin{table}[th]
\caption{Sample $\delta$ values for various digest length and LDs.}\label{tab:delta_values}
\centering\small
\begin{tabular}{lllll}
\toprule
  & $len(dig_A)$ & $len(dig_B)$ & $LD(dig_A,dig_B)$ & $delta$ \\
\midrule
1 & 700          & 700          & 0                 & 1.000   \\
2 & 700          & 700          & 10                & 0.986   \\
3 & 700          & 350          & 400               & 0.857   \\
4 & 700          & 100          & 600               & 1.000   \\ \hdashline
5 & 700          & 700          & 600               & 0.143   \\
6 & 700          & 350          & 650               & 0.143   \\
7 & 700          & 100          & 696               & 0.040   \\
8 & 700          & 200          & 700               & 0.000   \\ \hdashline
9 & 70'000       & 700          & 70'000            & 0.000   \\
10& 70'000       & 700          & 69'650            & 0.500   \\
11& 70'000       & 700          & 69'300            & 1.000   \\
\bottomrule
\end{tabular}
\end{table}

There are several possible solutions: 
(i) define a multiplier so that both digest lengths have to be in this range from each other (e.g., factor 10); 
(ii) adjust the equation to consider this behavior (e.g., using $R_{Random}$); or 
(iii) require that sequences share a common substring of a certain length (e.g., if $dig_B$ is a substring of $dig_A$, they are likely related).

\subsection[Assessment significance score]{Assessment significance score ($\delta$)}
To assess $\delta$'s raison d'\^etre, we performed two test runs: (1) do an all-against-all comparison of 72 unrelated files and (2) finding 10 originals in the full dataset.

\paragraph{Test1} To create the 72-unrelated-files dataset, we used the large set, randomly selected 72 files and shrank them to 30\,KB by removing text from the beginning and the end (i.e., deleting potential header/footer information). We then calculated the LD between all documents and sorted the 2556-pairs according to their LD ranging from to 22'255 to 25'762. Lastly, we selected some pairs with a low distance and analyzed them manually to ensure they are different. While further testing is required, we believe these still significant overlaps originate from the many leading spaces in the documents (see example in Fig.~\ref{fig:book_leading_sapce})\footnote{For one pair, we removed the double-space which increased the LD by about 2200.}. 

\begin{figure}[th]
\caption{Extract from document 13649.txt showing the leading spaces which cause relatively low LD's.}\label{fig:book_leading_sapce}
\centering \scriptsize
\begin{Verbatim}[frame=single]
    Of the Hills of the Chankly Bore,--

    Then, through the vast and gloomy dark
    There moves what seems a fiery spark,--
        A lonely spark with silvery rays
        Piercing the coal-black night,--
        A Meteor strange and bright:
    Hither and thither the vision strays,
        A single lurid light.
\end{Verbatim}
\end{figure}

Next, we ran our algorithm ($C=51$, $N=11$) and computed the significance score $\delta$ for each pair. The summarized results are shown in Table~\ref{tab:delta_unrelated_files}. As expected, we obtain low scores for all matches where only one match is above 0.12. To conclude, $\delta$ works for dissimilar documents that are of same/similar size.

\begin{table}[ht]
    \caption{Summary of the $\delta$ values for unrelated files. Right-half of the table shows accumulated results where all matches where almost all matches where lower than 0.12.}\label{tab:delta_unrelated_files}
    \centering
    \resizebox{\columnwidth}{!}{%
    \begin{tabular}{rrr|rrrrr}
        \toprule
         Min    & Max   & Avg       & $\le 0.03$       & $\le 0.06$    & $\le 0.09$    & $\le 0.12$    & $\le 0.15$   \\
         0.025  & 0.122 & 0.058     & 19               & 1470          & 2479          & 2555          & 2556         \\
         \bottomrule
    \end{tabular}
    }
\end{table}

\paragraph{Test 2} Our second test focused on identifying 10 of these unrelated documents in the whole dataset. Therefore, we created two signature files and used the source-destination mode of our application. As the dataset is rather large, we set $C=101$ ($N$ remained 11). Additionally, we manipulated the application to only consider files that are up to 10 times larger than the source file (if larger, $\delta$ is set to zero). The distribution of the $\delta$ score is shown in Table~\ref{tab:distribution_delta_score_matches} where the top rows indicate the $\delta$-score and the bottom rows the absolute number of pairs.

\begin{table}[ht]
    \caption{Distribution of the $\delta$ score from comparing 10 documents against the dataset (18'196).}\label{tab:distribution_delta_score_matches}
    \centering
    \resizebox{\columnwidth}{!}{%
    \begin{tabular}{rrrrr}
        \toprule
      $\delta \le 0.1$   & $0.1<\delta\le0.2$   & $0.2<\delta\le0.3$    & $0.3<\delta\le0.4$    & $0.4<\delta\le0.5$    \\
        93'359           & 9357                 & 18'193                & 25'743                & 24'298  \\
     \midrule
     $0.5<\delta\le0.6$     & $0.6<\delta\le0.7$    & $0.7<\delta\le0.8$    & $0.8<\delta\le0.9$        & $0.9<\delta$\\
     10'998                 &2                      &0                      &0                          & 10\\
     \bottomrule
    \end{tabular}
    }
\end{table}

As indicated by the table, the 10 matches were clearly identified with a gap to any other matches. 
We manually investigated the two matches for $0.6 < \delta \le 0.7$ and did not see any similarity. However, both files were close to 300\,KB which is maximum difference in size to be considered and again highlights the weakness of the current $\delta$-formula. 


\section{Related work}\label{sec:rel_work}
Given the sheer amount of work that has been done in terms of string similarity and approximation,  we only included a summary where we decided to split the section into two domains: string similarity / approximate string matching and approximate matching (as defined from the digital forensics community).

\paragraph{String similarity} String similarity metrics have a long history and may different techniques exist. \cite{gang2020string} proposes to divide them in four categories: 
(1) character-based / edit distance (examples are Hamming Distance, Levenshtein Distance, Damerau-Levenshtein Distance or Jaro Distance);
(2) Sequence-based (examples are Longest Common Subseqeunce, Longest Common Substring, or Geshtalt Pattern Matching);
(3) Token-based (examples are Q-Grams, Overlap Coefficient or Jakkard Similarity); and
(4) Phonetic-sensitive (examples are Smith-Waterman or Editex). 
Our work falls into the first category, which is also known as the string-to-string correction problem \citep{wagner1974string}. Many of these techniques have been analyzed in detail, optimized algorithms have been published and improvements have been presented. A thorough analysis of work related to the Levenshtein Distance has been done by \cite{navarro2001guided} who compared the performance of different algorithms doing a number of experiments. However, most works focus on improving and not on estimating. For instance, \cite{ukkonen1985algorithms} proposes an improvement that runs in $O(s\cdot min(len(A),len(B))$ where $s$ is the edit distance of the two strings compared. Thus, the algorithm runs in approximately linear runtime for highly similar strings.

\paragraph{Approximate matching (AM)} From a digital forensics perspective, these algorithms are interesting but despite their improvements, too slow for practical usage on large amounts of data. Therefore, the community invented its own AM (a.k.a.~fuzzy hashing for similarity hashing) which later has been summarized by the National Institute of Standards and Technologies in SP800-168 with the goal to ``to provide a definition and terminology to describe approximate matching in order to promote discussion, research, tool development and tool acquisition'' \citep{breitinger2014approximate}.
The research in this domain evolves around developing novel algorithms (e.g.,  \cite{kornblum2006identifying,oliver2013tlsh,chang2019fbhash}), comparing algorithms (e.g., \cite{roussev2011evaluation}) or improving/assessing existing implementations (e.g., \cite{baier2011security}). 
Discussing all existing works is beyond the scope of this article but a somewhat up to date summary is provided by \cite{harichandran2016bytewise} as well as \cite{martin2021bringing} who published a categorization of AM algorithms including a discussion of various algorithms.
A point that is raised in many publications is the similarity score (i.e., the result when comparing two fingerprints) which is only an indicator of similarity but the similarity is not clearly defined, i.e., a similarity score of 50 does not meant that two inputs are 50 percent identical or that 50 paragraphs are identical. While this indicator (two inputs are similar with a high certainty) is sufficient for some use cases, other scenarios may require a more precise result. One example would be checking for plagiarisms as often done by university during online submissions. 

\cite{breitinger2014similarity} published an application named \texttt{saHash} (statistical analysis hashing) which is based on four sub-hash functions, each  creating its own sub-hash value. The final, fixed-length hash is then the concatenation of all sub-hashes. While the authors claim that their implementation is almost as fast as SHA-1, it only ``enables the detection of  `small' changes between files – up to several hundred Levenshtein operations''.

\section{Conclusions}\label{sec:conclusion}
The Levenshtein Distance (LD) is a well-known distance metric to describe the dissimilarity between two strings / documents. While it works well for short sequences, it requires a quadratic runtime and thus is impractical for long sequences (e.g., hundreds of kilobytes). In this article we presented a heuristic that allows estimating the LD also for larger sequences. The algorithm works by first compressing a given document into a signature.
and then comparing signatures against each other to estimate the LD. 
Consequently, it is not necessary to possess the original documents to calculate their LD.
To assess the quality of our algorithm, we performed several experiments and show that our estimation is significantly faster and, depending on the modifications done to a source element, obtains accurate estimations (e.g., the experiments showed that estimations are very accurate for dissimilar documents of the same size as well as that `deletions' are handled well.

\paragraph{Future Research}
While the effectiveness of the heuristic seems clear, there are several areas that need further investigation, especially the following four:
(1) As outlined in the article, the calculation of $\delta$ is not perfect and we had to introduce a multiplier that documents have to be of similar size (e.g., within factor 10). It requires more experiments to see what mechanism provides the best results from a digital forensic perspective. 
(2) 
Documents can have a large LD despite sharing substantial blocks of identical or nearly identical text.  One way to detect pairs that are related in this way is to divide large documents into uniform sized blocks and estimate the LD's of all pairwise combinations of blocks from the two originals. This would allow relatively small segments, that are similar, to stand out against a background of unrelated text. 
(3) 
File signatures are like thumbnail images, with the differences between a pair of signatures providing a coarse map of the differences between the corresponding originals. A better understanding how to characterize and quantify different kinds of similarity, and understand the limitations of this technique would make this property more useful.
(4) 
All experiments have been on books in English that have been coarsely mutilated, usually with many blocks of lines either being deleted or duplicated.  A more systematic survey of the quality of estimates based on files with quantified degrees and kinds of differences would be useful. The behavior on particular types of data would also be useful. This would be of particular interest in applying the algorithm to binary data such as video or compiled computer programs.

\bibliography{bibl/literature.bib}







\appendix

\section{eLD Java Implementation}\label{app:eLD_java}
The implementation of our algorithm is written entirely in Java 11 and compiled using Maven 3.6.3 and OpenJDK 64-Bit Java version: 11.0.11. The source code is generic Java using no language features not found in any standard imperative computer language. To keep descriptions of performance simple, the current version is single-threaded only. It is open-source and can be downloaded here: \url{https://coatespt.github.io/Fast-Levenshtein-Distance/}. 
The application supports command line arguments or the usage of a configuration file. To generate signatures run
\begin{command}
java -jar Fast-LD.jar -p config\_gen.properties
\end{command}
\noindent and to compare the files-folder against an existing set of signatures (\texttt{sigs.csv}) run
\begin{command}
java -jar Fast-LD.jar -p config\_cmp.properties
\end{command}

In these examples, our Fast-LD implementation utilizes configuration files (CLI arguments overwrite the config) which is listed in Fig~\ref{fig:conf_cmp_file}. 
Parameters $C$, $N$, and $CH$ (Alphabet) have been a explained in Sec.~\ref{sec:comp_algo_des}; 
\texttt{f} points to a CSV file whose content is a list of files to be processes (one per line); 
\texttt{ld=true} indicates comparison mode (Levenshtein Distance Estimation); and 
\texttt{ft} is the signature database in the CSV format.
In order to generate signatures (no comparison), one only has to set \texttt{ld=false} (for better readability, \texttt{ft} can be removed as it is not needed / will be ignored).

\begin{figure}[th]
\caption{Content of the configuration file \texttt{config\_cmp.properties}; the alphabet has been shortened for better readability.}\label{fig:conf_cmp_file}
\scriptsize
\begin{Verbatim}[frame=single]
c = 251 
n = 11
f = input.csv
ch = abcdefghijklnmnoprstuv[...]234567890
ld = true
ft = sigs.csv
\end{Verbatim}
\end{figure}


\end{document}